\documentclass[aps,showpacs,epsfig,twocolumn]{revtex4}
\usepackage{epsfig}
\usepackage{color}
\usepackage{bbm}
\usepackage{epsfig}
\usepackage{color}
\usepackage{amsmath}
\usepackage{amssymb}
\usepackage{bbm}

\newcommand{\be}{\begin{equation}}
\newcommand{\ee}{\end{equation}}
\newcommand{\bea}{\begin{eqnarray}}
\newcommand{\eea}{\end{eqnarray}}

\begin{document}

\title{\bf\Large {Leggett's Modes in Magnetic Systems with Jahn-Teller distortion}}

\author{Naoum Karchev }

\affiliation{Department of Physics, University of Sofia, 1126 Sofia, Bulgaria }

\begin{abstract}
Leggett's mode is a collective excitation corresponding to the oscillation of the  relative phase of the order parameters  in a two band superconductor, with frequency proportional to interband coupling. We report on the existence of modes, similar to Leggett's mode, in magnetic systems with Jahn-Teller distortion. The minimal Kugel-Khomskii model, which describes simultaneously both the spin and the orbital order, is studied. The dynamical degrees of freedom are spin-$s$ operators of localized spins and pseudospin-$\tau$ operators, which respond to the orbital degeneracy and satisfy the similar commutation relation with
those of the spin operators. In the case of "G-type antiferro" spin and pseudospin order the system possesses two antiferromagnetic magnons with equal spin-wave velocities and two Leggett's modes with equal gaps proportional to the square root of the spin-pseudospin interaction constant. In the case of "ferro" spin and pseudospin order the system possesses one ferromagnetic magnon and one Leggett's mode with gap proportional to the spin-pseudospin interaction constant. We conclude that Leggett's modes, in the spectrum of the magnetic systems with Jahn-Teller distortion, are generic feature of these systems.

\end{abstract}

\maketitle

\emph{Introduction}-The spontaneous breaking of a continuous symmetry is accompanied with long range excitation known as Goldstone mode. In systems with two or more order parameters the Goldstone boson is supplemented by excitation which, in some sense, is orthogonal to it and has a mass proportional to the constant of interaction between different order parameters.

In the theory of superconductivity the phase of the order parameter is a massless excitation known as Anderson-Bogoliubov-Goldstone (ABG) mode \cite{Anderson58,Bogoliubov59}. In two band superconductor the (ABG) mode is a combination of the phases of the order parameters. It is complemented by a mode associated with the relative phases oscillation with frequency proportional to interband coupling \cite{Leggett66}. The Leggett's mode was observed in $MgB_2$ superconductor with Raman spectroscopy \cite{Blumberg07}. A novel peak in the one of the scattering channels is observed. The authors assign this feature to the Leggett's mode. The measured mass is in accordance with theoretically predicted one \cite{Sharapov02}. In superconductors with three and more bands there are multiple Leggett's modes classified by multiple interband couplings \cite{Ota11}.

An analogous Leggett's mode is theoretically predicted in superconductor with mixed-symmetry order parameter generated in an external magnetic field \cite{Balatsky00}. The oscillations of the relative phase (Leggett's mode) are with frequency proportional to the magnetic field.

An important class of magnetic materials are compounds in which the state of magnetic ions is characterized by orbital as well as spin degeneracy.
According to the Jahn-Teller theorem \cite{JT}, an atom configuration in which orbital degeneracy is realized is unstable. The symmetry is
lowered and the degeneracy is lifted, corresponding to ordering of the orbitals.

In a mathematical description of a two orbital system it is convenient to introduce pseudospin $\tau=1/2$ associated with the two bands, in a way that the one of the band corresponds to the value $\tau^z=1/2$, while the other one corresponds to the value $\tau^z=-1/2$. To model the electron-phonon coupling it is convenient to introduce three-component pseudospin operators $T_{\alpha}$ which satisfy the similar commutation relation with those of the spin operator, i.e., $[T_{\alpha}, T_{\beta}]=i\varepsilon_{\alpha\beta\gamma}T_{\gamma}$. Eliminating the phonons from the theory one obtains an effective theory with Hamiltonian which can be written in a form of Heisenberg (or Ising) Hamiltonian for pseudospins. There is an interaction between the spin and pseudospin of the ions. Collecting all terms including spin exchange ones we obtain the effective  Kugel-Khomskii  model\cite{KK80,KK82}.

The dynamical degrees of freedom are spin-$s$ operators of localized spins and pseudospin-$\tau$ operators, which respond to the orbital degeneracy and satisfy the similar commutation relation with
those of the spin operators. In the case of "G-type antiferro" spin and pseudospin order the system possesses two antiferromagnetic magnons with equal spin-wave velocities and two Leggett's modes with equal gaps proportional to the square root of the spin-pseudospin interaction constant. In the case of "ferro" spin and pseudospin order the system possesses one ferromagnetic magnon and one Leggett's mode with gap proportional to the spin-pseudospin interaction constant.

\emph{Kugel-Khomskii  model}- The Hamiltonian of the minimal model is
\bea
 h & = & J^s \sum\limits_{  \langle  ij  \rangle  } {{\bf S}_i \cdot {\bf S}_j}+J^p \sum\limits_{  \langle  ij  \rangle  } {{\bf T}_i
\cdot {\bf T}_j}\nonumber \\
& - & J \sum\limits_{  \langle  ij  \rangle  } ({\bf S}_i\cdot {\bf S}_j)({\bf T}_i\cdot {\bf T}_j)\label{LMJTh},
\eea
where ${\bf S}_i$ is spin-$s$ operator, ${\bf T}_i$ is pseudospin-$\tau$ operator, $J^s$ is spin exchange constant, $J^p$ is pseudospin exchange constant and $J$ is spin-pseudospin interaction constant.  The sums are over all sites of a three-dimensional cubic lattice, and $\langle i,j\rangle$ denotes the sum over the nearest neighbors.

We consider a system with "antiferro" spin and pseudospin order. All constants in the Hamiltonian Eq.(\ref{LMJTh}) are positive $(J^s>0,\,\, J^p>0,\,\, J>0)$. To proceed we treat the spin-pseudospin interaction in the Hartree-Fock approximation. To this end one represents the term of interaction in the form
\bea
& & ({\bf S}_i\cdot {\bf S}_j)({\bf T}_i\cdot {\bf T}_j)_{HF} = -\langle S_i^{\alpha} T_i^{\beta} \rangle \langle S_j^{\alpha} T_j^{\beta} \rangle \nonumber \\
& &  + \langle S_i^{\alpha} T_i^{\beta} \rangle S_j^{\alpha} T_j^{\beta} + \langle S_j^{\alpha} T_j^{\beta} \rangle S_i^{\alpha} T_i^{\beta}
\label{LMJT-HF}
\eea
with
$\langle S_i^{\alpha} T_i^{\beta} \rangle = \frac {\delta^{\alpha \beta}}{3} s \tau v$, where $v$ is the Hartree-Fock  parameter,  to be determined self-consistently.
The Hamiltonian Eq.(\ref{LMJTh}) in the Hartree-Fock approximation reads
\bea
h_{HF} & = & J^s \sum\limits_{  \langle  ij  \rangle  } {{\bf S}_i \cdot {\bf S}_j}+J^p \sum\limits_{  \langle  ij  \rangle  } {{\bf T}_i
\cdot {\bf T}_j}\nonumber \\
& - & 2 s v \tau J \sum\limits_{i} {\bf S}_i\cdot {\bf T}_i + J s^2 \tau^2 v^2 N\label{LMJTh2}
\eea
where $N$ is the number of the lattice's sites. Equation (\ref{LMJTh2}) shows that the Hartree-Fock parameter renormalizes
the spin-pseudospin interaction constant $J_r=2 s \tau v J$.

To study a theory with Hamiltonian Eq.(\ref{LMJTh2}) it is convenient to introduce Holstein-Primakoff representation for the spin ${\bf S}_j(a^+,a)$
and pseudospin ${\bf T}_j(b^+,b)$ operators
\begin{eqnarray}\label{LMJT-HP}
S_j^+ & = & S_j^1+iS_j^2 \nonumber \\
& = & \cos^2\frac {\theta_j}{2} \sqrt{2s-a_j^+a_j}\, a_j - \sin^2\frac {\theta_j}{2} a_j^+ \sqrt{2s-a_j^+a_j}\nonumber \\
S_j^- & = & S_j^1-iS_j^2  \\
& = & \cos^2\frac {\theta_j}{2} a_j^+ \sqrt{2s-a_j^+a_j} - \sin^2\frac {\theta_i}{2} \sqrt{2s-a_j^+a_j}\,  a_j \nonumber \\
S_j^3 & = & \cos\theta_j (s-a_j^+ a_j), \nonumber
\end{eqnarray}
where  $\theta_j={\bf Q}\cdot {\bf r}_j$ and ${\bf Q}=(\pi,\pi,\pi)$ is the antiferromagnetic wave vector.
The representation for the pseudospin operators ${\bf T}_j(b^+,b)$ is obtained from Eq.(\ref{LMJT-HP}) replacing Bose operators $(a^+,a)$ with Bose operators $(b^+,b)$ and spin-$s$ with pseudospin-$\tau$.
In terms of the Bose fields and
keeping only the quadratic terms, the effective Hamiltonian Eq.(\ref{LMJTh2}) adopts the form
\bea\label{LMJT-h3}
h_{HF} & = & N s^2 \tau^2 J (v-1)^2- N s^2 \tau^2 J + h_q \\
h_q & = & s J^s \sum\limits_{  \langle  ij  \rangle  }(a^+_ia_i + a^+_ja_j -a^+_ia^+_j-a_ja_i) \nonumber \\
& + & \tau J^p \sum\limits_{  \langle  ij  \rangle  }(b^+_ib_i + b^+_jb_j -b^+_ib^+_j-b_jb_i) \nonumber \\
& + & J_r \sum\limits_{i}[\sqrt{s\tau}(a^+_ib_i+b^+_ia_i)-\tau a^+_ia_i-sb^+_ib_i]. \nonumber
\eea

To proceed one rewrites the Hamiltonian $h_q$ in the momentum space representation:
\bea \label{LMJT-h4}
 h_q & = & \sum\limits_{k\in B}\left [\varepsilon^a\,a_k^+a_k\,+\,\varepsilon^b\,b_k^+b_k\,-
 \,\gamma \left (a_k^+b_k + b^+_k a_k \right )\,\right. \nonumber \\
 & - & \left. \gamma_k^a(a^+_ka^+_{-k}+a_{-k}a_k)-\gamma_k^b(b^+_kb^+_{-k}+b_{-k}b_k)\right ],
\eea
where the wave vector $k$ runs over the  first Brillouin zone and the dispersions are given by equalities
\bea \label{LMJT-h5}
\varepsilon^a & = & 6s J^s + \tau J_r , \quad \varepsilon^b=6\tau J^p +s J_r, \quad  \gamma=J_r\sqrt{s \tau}, \nonumber \\
\gamma^a_k & = & s J^s (\cos k_x +\cos k_y +\cos k_z), \\
\gamma^b_k & = & \tau J^p (\cos k_x +\cos k_y +\cos k_z).\nonumber
\eea

To diagonalize the Hamiltonian we introduce new Bose fields
$\alpha_k,\,\alpha_k^+,\,\beta_k,\,\beta_k^+$ by means of the Bogoliubov
transformation. The technique for diagonalization developed in \cite{White65} is used. The details are given in the supplementary materials \cite{LMJT-SM}.
The transformed Hamiltonian has the form
\be
\label{LMJTh5} h_q = \sum\limits_{k\in B}\left
(E^{\alpha}_k\,\alpha_k^+\alpha_k\,+\,E^{\beta}_k\,\beta_k^+\beta_k\,+\,E^0_k\right),
\ee
with dispersions
\bea  \label{LMJT_E}
& & E^{\alpha}_k\,=\,\sqrt{\frac
12\,\left (A_k+B_k-\sqrt{(A_k-B_k)^2+4 D_k} \right)} \nonumber \\
\\
& & E^{\beta}_k\,=\,\sqrt{\frac
12\,\left (A_k+B_k+\sqrt{(A_k-B_k)^2+4 D_k} \right)}, \nonumber  \eea
where
\bea\label{LMJTE2}
A_k & = & (\varepsilon^a)^2 + \gamma^2 - 4 (\gamma^a_k)^2 \nonumber \\
B_k & = & (\varepsilon^b)^2 + \gamma^2 - 4 (\gamma^b_k)^2  \\
D_k & = & \gamma^2\left [\left (\varepsilon^a-\varepsilon^b \right)^2-4\left (\gamma^a_k  - \gamma^b_k \right )^2\right ].\nonumber
\eea
and $E^{0}_k$ is the vacuum energy \cite{LMJT-SM}.

To determine self-consistently the Hartree-Fock parameter we calculate the free-energy of the system as a function of the parameter $v$
\bea\label{LMJT F}
& &  F(v) = J s^2 \tau^2 (v-1)^2-J s^2 \tau^2 +\frac 1N \sum\limits_{k\in B}E^0_k \\
& + & \frac {T}{N}\sum\limits_{k\in B}[\ln (1-\exp[-E_k^{\alpha}/T]) + \ln (1-\exp[-E_k^{\beta}/T])].\nonumber
\eea
The physical value of the parameter is the value at which the free-energy has a minimum.
The dimensionless free-energies $F(v)/J$, as a function of the Hartree-Fock parameter $v$ at zero temperature, are depicted in figure (1).
\begin{figure}[!ht]
\epsfxsize=\linewidth
\epsfbox{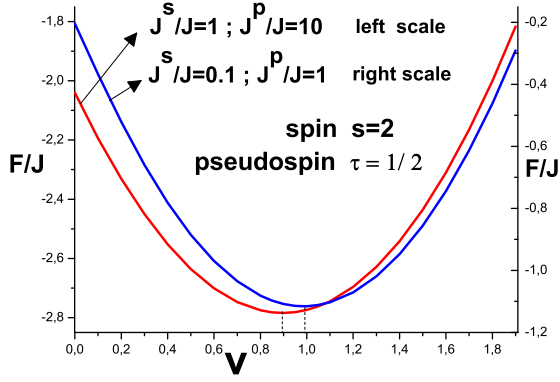} \caption{(Color online)\,The dimensionless free-energies $F(v)/J$, as a function of the Hartree-Fock parameter $v$, for a spin $s=2$ and pseudospin $\tau=1/2$ system with parameters $J^s/J=1$ and $J^p/J=10$-left scale (red line) and  for $J^s/J=0.1$ and $J^p/J=1$-right scale (blue line). The Hartree-Fock parameter for the first one is $v=0.894$, while for the second system it is $v=0.987$. }
\label{LMJT F}
\end{figure}
For a spin $s=2$ and pseudospin $\tau=1/2$ system with parameters $J^s/J=1$ and $J^p/J=10$ (left scale-red line), one obtains $v=0.894$. For $J^s/J=0.1$ and $J^p/J=1$ (right scale-blue line) the Hartree-Fock parameter is $v=0.987$.

The equations (\ref{LMJT_E}) and (\ref{LMJTE2}) show that dispersions $E^{\alpha}_k$ and $E^{\beta}_k$ depend on the wave vector ${\bf k}$ through the expression $\varepsilon_k=\cos k_x +\cos k_y +\cos k_z$. It is convenient to draw these energies as functions of $\varepsilon_k$. The figure (2) shows the functions $E^{\alpha}(\varepsilon_k)$ and $E^{\beta}(\varepsilon_k)$ for a system with parameters $s=2$, $\tau=1/2$, $J^s/J=1, J^p/J=10$ and $v=0.894$.
\begin{figure}[!ht]
\epsfxsize=\linewidth
\epsfbox{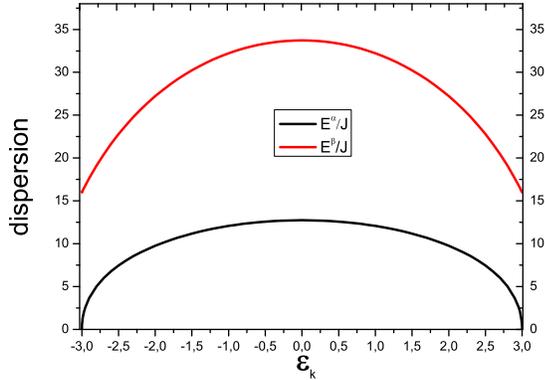} \caption{(Color online)\,The dimensionless energies $E^{\alpha}$ (black line) and $E^{\beta}$ (red line) as functions of $\varepsilon_k$ $(-3<\varepsilon_k<3)$. }
\label{TN-tU}
\end{figure}

It is evident that $E^{\alpha}_k$ is equal to zero at ${\bf k}=(0,0,0)$ and ${\bf k}_{\pi}=(\pm \pi,\pm \pi,\pm \pi)$. Therefor, the two branches of the $\alpha_k$-boson describe the two long-range excitations (magnons) in the system \cite{LMJTnote}. Near these vectors the dispersion adopts the form $E_k^{\alpha}\propto c_s |\textbf{k}|$ and $E_k^{\alpha}\propto c_s |\textbf{k}-\textbf{k}_{\pi}|$,
with equal spin-wave velocity $c_s$.

The energy $E^{\beta}_k$ has a minimum at
${\bf k}=(0,0,0)$ and ${\bf k}_{\pi}=(\pm \pi,\pm \pi,\pm \pi)$ and
\bea\label{LMJTgap}
E^{\beta}_0 & = & E^{\beta}_{k_{\pi}}=\Delta^L \\
\Delta^L & = & \sqrt{2s\tau v J}\sqrt{12s\tau (J^s+J^p)\,+\,2s\tau v J (s+\tau)^2}\nonumber .
\eea
The gap is proportional to $\sqrt{J}$ which means that the two branches of the $\beta_k$-boson describe the two Leggett's modes in the system with equal gaps.

To illustrate the relationship between the geometry of the magnetic order and the nature of Leggett's mode we consider a system with "ferro" spin and pseudospin order. The Hamiltonian of the system Eq.(\ref{LMJTh}) has negative spin-exchange $J^s<0$ and pseudospin-exchange $J^p<0$ constants. We obtain, in the same way, the Hartree-Fock Hamiltonian Eq.(\ref{LMJTh2}) and use the Holstein-Primakoff representation Eq.(\ref{LMJT-HP})with feromagnetic wave vector ${\bf Q}=(0,0,0)$. In terms of the Bose fields and keeping only the quadratic terms, the effective Hamiltonian Eq.(\ref{LMJTh2}) adopts the form
\bea\label{LMJT-h-f}
h_{HF} & = & N s^2 \tau^2 J (v-1)^2- N s^2 \tau^2 J + h_q \\
h_q & = & s |J^s| \sum\limits_{  \langle  ij  \rangle  }(a^+_ia_i + a^+_ja_j -a^+_ia_j-a^+_ja_i) \nonumber \\
& + & \tau |J^p|\sum\limits_{  \langle  ij  \rangle  }(b^+_ib_i + b^+_jb_j -b^+_ib_j-b^+_jb_i) \nonumber \\
& + & J_r \sum\limits_{i}[\sqrt{s\tau}(a^+_ib_i+b^+_ia_i)-\tau a^+_ia_i-sb^+_ib_i]. \nonumber
\eea
We rewrite the Hamiltonian $h_q$ in the momentum space representation:
\bea \label{LMJT-h-f2}
 h_q & = & \sum\limits_{k\in B}\left [\varepsilon^a_k\,a_k^+a_k\,+\,\varepsilon^b_k\,b_k^+b_k\,-
 \,\gamma \left (a_k^+b_k + b^+_k a_k \right )\,\right]\nonumber \\
 \varepsilon^a_k\, & = & 2s |J^s| (3-\cos k_x -\cos k_y -\cos k_z)\,+\,\tau J_r\nonumber \\
 \varepsilon^b_k\, & = & 2\tau |J^p| (3-\cos k_x -\cos k_y -\cos k_z)\,+\,s J_r\nonumber \\
 \gamma & = & \sqrt{s \tau} J_r
\eea
To diagonalize the Hamiltonian Eq.(\ref{LMJT-h-f2}) we introduce new Bose fields
$\alpha_k,\,\alpha_k^+,\,\beta_k,\,\beta_k^+$ by means of rotation. The transformed Hamiltonian has the form
\be
\label{LMJTf-h3} h_q = \sum\limits_{k\in B}\left
(E^{\alpha}_k\,\alpha_k^+\alpha_k\,+\,E^{\beta}_k\,\beta_k^+\beta_k\right),
\ee
with dispersions
\bea  \label{LMJT-f-E}
& & E^{\alpha}_k\,=\,\frac
12\,\left [\varepsilon^a_k+\varepsilon^b_k- \sqrt{(\varepsilon^a_k-\varepsilon^b_k)^2+4\gamma^2}\right] \nonumber \\
\\
& & E^{\beta}_k\,=\,\frac
12\,\left [\varepsilon^a_k+\varepsilon^b_k +\sqrt{(\varepsilon^a_k-\varepsilon^b_k)^2+4\gamma^2}\right] \nonumber  \eea

The free energy of the system is
\bea\label{LMJT f-F}
& &  F(v) = J s^2 \tau^2 (v-1)^2-J s^2 \tau^2 \\
& + & \frac {T}{N}\sum\limits_{k\in B}[\ln (1-\exp[-E_k^{\alpha}/T]) + \ln (1-\exp[-E_k^{\beta}/T])].\nonumber
\eea
One obtains that at zero temperature the physical value of the Hartree-Fock parameter is $v=1$ for all values of the parameters.

It follows from equations (\ref{LMJT-h-f2}) and (\ref{LMJT-f-E}) that $E^{\alpha}_0=0$ and near the zero wave vector the $\alpha$-boson has a ferromagnetic dispersion $E^{\alpha}_k=\propto \rho {\bf k}^2$ with spin-stiffness $\varrho=(s^2|J^s|+\tau^2|J^p|)/(s+\tau)$. On the other hand, $\beta$-boson is gapped excitation with gap
\be\label{LMJT-f-gap}
\Delta^L=E^{\beta}_0=2s\tau (s+\tau)J,\ee
where $J$ is the spin-pseudospin interaction constant. This means that $\beta$-boson is the Leggett's mode in the system.

\emph{Kugel-Khomskii  model with Ising pseudospin anisotropy}- The Hamiltonian of the system is
\be\label{LMJT-Ising1}
\hat{h}=h-\Delta J \sum\limits_{  \langle  ij  \rangle  } {T_i^z
\cdot T_j^z}\ee
where $h$ is the Hamiltonian Eq.\eqref{LMJTh}, and $\Delta J>0$ is the anisotropy parameter. While the $SU(2)$ pseudospin symmetry is broken, one can use the representations \eqref{LMJT-HP} for the spin ${\bf S}_j(a^+,a)$ and pseudospin ${\bf T}_j(b^+,b)$ operators, and following the same technique of calculation to obtain the spectrum of a system with negative (ferro) exchange constants $J^s<0$ and  $J^p<0$(see Eqs.\eqref{LMJTf-h3})
\bea\label{LMJT-Ising2}
\hat{h}_q & = &  \sum\limits_{k\in B}\left
(\hat{E}^{\alpha}_k\,\alpha_k^+\alpha_k\,+\,\hat{E}^{\beta}_k\,\beta_k^+\beta_k\right) \\
 \hat{E}^{\alpha}_k & = & \frac
12\,\left [\varepsilon^a_k+\hat{\varepsilon}^b_k- \sqrt{(\varepsilon^a_k-\hat{\varepsilon}^b_k)^2+4\gamma^2}\right] \nonumber \\
\hat{E}^{\beta}_k & = & \frac
12\,\left [\varepsilon^a_k+\hat{\varepsilon}^b_k +\sqrt{(\varepsilon^a_k-\hat{\varepsilon}^b_k)^2+4\gamma^2}\right]. \nonumber  \eea
In Eqs.\eqref{LMJT-Ising2} $\hat{\varepsilon}^b_k\,=\,\varepsilon^b_k+6\tau\Delta J$ with $\varepsilon^a_k, \varepsilon^b_k$ and $\gamma$ from Eqs.\eqref{LMJT-h-f2} .
The energies  $\hat{E}^{\alpha}_k$ and $\hat{E}^{\beta}_k$ have a minimum at
${\bf k}=(0,0,0)$
\bea\label{LMJT-Ising3}
\hat{E}^{\alpha}_0 = \frac
12\,\left [\Delta^L+6\tau\Delta J- \sqrt{(\Delta^L+6\tau\Delta J)^2-24\tau^2J_r\Delta J}\right] \nonumber \\
\\
\hat{E}^{\beta}_0 = \frac
12\,\left [\Delta^L+6\tau\Delta J+\sqrt{(\Delta^L+6\tau\Delta J)^2-24\tau^2J_r\Delta J}\right], \nonumber  \eea
where  $\Delta^L$ is the Leggett's gap in an isotropic system Eq.\eqref{LMJT-f-gap}.
Both dispersions have a gap but we can identify the Legget's mode as an excitation with the larger one $\hat{E}^{\beta}_0>\hat{E}^{\alpha}_0 $. In the limit of small anisotropy the leading contribution of the anisotropy parameter $\Delta J$ to the dispersions is

\bea\label{LMJT-Ising4}
& & \hat{E}^{\alpha}_0 \approx  \frac {6\tau^2}{s+\tau} \Delta J \nonumber \\
\\
& & \hat{E}^{\beta}_0 \approx \Delta^L + \frac {6\tau s}{s+\tau} \Delta J. \nonumber  \eea
Eqs.\eqref{LMJT-Ising4} show that the gap of the $\alpha$ excitations is due to the pseudospin anisotropy, while the gap of the Leggett's mode is a sum of the gap due to the anisotropy and Leggett's gap.

\emph{Summary}-In the present paper we have studied theoretically the existence of Leggett's modes in magnetic systems with Jahn-Teller distortion.
It is theoretically predicted that a system with "G-type-antiferro" spin and pseudospin order possesses two antiferromagnetic magnons with equal spin-wave velocities and two Leggett's modes with equal gaps proportional to the square root of the spin-pseudospin interaction constant.

A prominent example of magnetic system with Jahn-Teller distortion is the $LaMnO_3$ compound with "perovskite" structure. The magnetic $Mn$ ion has an incomplete 3d-shell. The trivalent $Mn^{3+}$ ion has four electrons. It is surrounded by six oxygen $O^{2-}$ ions which form an octahedral structure. The crystal field of these ligands results in a particular splitting of the five d-orbitals into well separated in energy two groups: the $e_g$ and $t_{2g}$ states. The $t_{2g}$ sector forms a triplet, and the $e_g$ one forms a doublet. The triplet state is lower and three of the d-electrons occupy $t_{2g}$ bands, while the last one occupies $x^2-y^2$ or $3z^2-r^2$ band of $e_g$ doublet \cite{Daggotto03}. At ambient conditions $LaMnO_3$ is a paramagnetic insulator. Below $T_N=140K$ the magnetic structure of the system is A-type antiferromagnetic \cite{Daggotto03}.
The strong distortion of the $MnO_6$ octahedra is the signature of the cooperative Jahn-Teller effect and orbital ordering\cite{Kanamori60,Goodenough70}. At $T_{JT}=750 K$ $LaMnO_3$ undergoes a structural phase transition above which the orbital ordering disappears\cite{RC98}. The C-type orbital structure in $LaMnO_3$ compound has been obtained experimentally by Y.Murakami et \emph{al.}\cite{Murakami98} and theoretically discussed in \cite{Hotta99}. The C-type orbital ordering means that if $e_g$ electron on site "i" occupies $x^2-y^2$ band the $e_g$ electron on nearest neighbor site in xy plane occupies $3z^2-r^2$ one, while along the z-direction the same orbital state repeats.

The model under consideration, in the present paper, do not match perfectly the $LaMnO_3$ compound. But magnon fluctuations in A-type, C-type and G-type antiferromagnets are identical, two Goldstone bosons with linear dispersion (see Supplemental material B). This is while we expect that theoretically predicted Laggett's modes are presented as well in the spectrum of the $LaMnO_3$ compounds.

There is an additional experimental evidence for this.
 Comparative Raman study of $LaMnO_3$ \cite{Iliev98} and $CaMnO_3$ \cite{Iliev02} shows that most intensive Raman line at $612 cm^{-1}$ in the spectra of   $LaMnO_3$ does not exist in the spectra of $CaMnO_3$. $LaMnO_3$ contains $Mn^{3+}$ ions with three $t_{2g}$ electrons and one $e_{g}$ electron which occupies $x^2-y^2$ or $3z^2-r^2$ band. This leads to Jahn-Teller effect in $LaMnO_3$. In $CaMnO_3$ the manganese is in $Mn^{4+}$ state with three $t_{2g}$ electrons and there is no Jahn-Teller effect. This pushes the authors to conclude that the most intensive Raman line at 612$cm^{-1}$ is a consequence of the Jahn-Teller effect \cite{Iliev03}.

Under the pressure \cite{Loa01} the Raman peak at $612\,\, cm^{-1}$ shifts towards higher energy and loses intensity with increasing pressure. There is strong indication that the Jahn-Teller effect and the concomitant orbital order are completely suppressed above $18\,\, GPa$. The Raman signal from Jahn-Teller distorted octahedra is still observed at $32\,\,GPa$ \cite{Mao11}.

The successful explanation of the extra Raman peak, in the two-band superconductor $MgB_2$, as a result of the Leggett's mode in the compound, inspires to assign the extra peak in $LaMnO_3$ to the Leggett's modes, theoretically predicted within the minimal Kugel-Khomskii model in the present paper.

There is a microscopical derivation of the effective Heisenberg model of A-type aniferromagnetism of $LaMnO_3$ compound \cite{Michev12}, but there is not such results  neither for C-type pseudo-spin antiferromagnetism nor for spin-pseudo-spin interaction. This does not permit a direct study of $LaMnO_3$ compound within Kugel-Khomskii theory.

The Hartree-Fock approximation (\ref{LMJT-HF}) is very important for our result. If one introduces the HF parameters $<S_i^\alpha S_j^\alpha>$ and
$<T_i^\beta T_j^\beta>$, to decompose the interaction, the resulting HF Hamiltonian is a sum of the Hamiltonian of the spin fluctuations and the Hamiltonian of the pseudo-spin fluctuations. As a result we have uncoupled Goldstone bosons which are transversal spin fluctuations and Goldstone bosons  which are transversal pseudo-spin fluctuations. Too many Goldstone bosons are not acceptable neither theoretically nor experimentally. The  Hartree-Fock approximation (\ref{LMJT-HF})
is the only way to mix the transversal spin and pseudo-spin fluctuations which leads to the correct spectrum.

The next step of our investigation is to understand the temperature dependence of the gap of the Leggett's mode. It follows from Eq.(\ref{LMJTgap})
that this dependence is through the Hartree-Fock parameter $v$. This parameter decreases when the temperature increases. To that purpose the gap  $\Delta^L(T)$ decreases with temperature increasing. The experimental measurement of the temperature dependence of the intensity of the Raman peak is very important for the understanding the relationship between Leggett's mode and Raman spectra.

Finally, we have studied a system with "ferro" spin and pseudospin order. This is the case when one orbital would be filled at each site. The spinel structures are the compounds , in which "ferro" deformations are favored \cite{KK82}. The temperature dependence of the gap is
\be
\Delta^L(T)/\Delta^L(0)=v(T),\ee
and one arrives at the conclusion that Langgett's gap decreases increasing the temperature in the same way as Hartree-Fock parameter does.

The author is grateful to Professor M. N. Iliev and Professor M. V. Abrashev for the useful discussions .

This work was partly supported by a Grant-in-Aid from Sofia University (2014).
\\ \\
\\ \\
\emph{Supplemental material A}

To diagonalize the Hamiltonian Eq.(\ref{LMJT-h4}) we introduce new Bose fields
$\alpha_k,\,\alpha_k^+,\,\beta_k,\,\beta_k^+$ by means of the Bogoliubov transformation:
\bea\label{LMJT-Btrans}
a_k & = & u^{11}_k \alpha_k + u^{12}_k \beta_k + v^{11}_k \alpha^+_{-k} +  v^{12}_k \beta^+_{-k} \nonumber \\
b_k & = & u^{21}_k \alpha_k + u^{22}_k \beta_k + v^{21}_k \alpha^+_{-k} +  v^{22}_k \beta^+_{-k}  \\
a^+_{-k} & = & \overline{v}^{11}_{-k}\, \alpha_k + \overline{v}^{12}_{-k}\, \beta_k + \overline{u}^{11}_{-k}\, \alpha^+_{-k} +  \overline{u}^{12}_{-k}\, \beta^+_{-k} \nonumber \\
b^+_{-k} & = & \overline{v}^{21}_{-k}\, \alpha_k + \overline{v}^{22}_{-k}\, \beta_k + \overline{u}^{21}_{-k}\, \alpha^+_{-k} +  \overline{u}^{22}_{-k}\, \beta^+_{-k}. \nonumber
\eea
All coefficients in the Hamiltonian Eq.(\ref{LMJT-h4}) are real and even functions of the wave vector $\bf k$. Therefore, we can consider a transformation Eq.(\ref{LMJT-Btrans}) with real parameters, which are even functions of the wave vector. One replaces the operators $a_k,\,b_k,\,a^+_{k},\,b^+_{k}$ in the Hamiltonian Eq.(\ref{LMJT-h4}) with operators $\alpha_k,\,\beta_k,\,\alpha^+_{k},\,\beta^+_{k}$, and imposes conditions the resulting Hamiltonian to be in a diagonal form Eq.(\ref{LMJTh5}).

To obtain the coefficients in the Bogoiubov transformation Eq.(\ref{LMJT-Btrans}), the new dispersions $E_k^{\alpha}, E_k^{\beta}$ and the vacuum energy $E_k^0$ we use the technique for diagonalization developed in \cite{White65}. Following this work one derives the inverse transformation
\bea\label{LMJT-INVtrans}
\alpha_k & = & u^{11}_k a_k + u^{21}_k b_k - v^{11}_k a^+_{-k} -  v^{21}_k b^+_{-k} \nonumber \\
\beta_k & = & u^{12}_k a_k + u^{22}_k b_k - v^{12}_k a^+_{-k} -  v^{22}_k b^+_{-k}  \\
\alpha^+_{-k} & = & -v^{11}_{k}\, a_k - v^{21}_{k}\, b_k + u^{11}_{k}\, a^+_{-k} +  u^{21}_{k}\, b^+_{-k} \nonumber \\
\beta^+_{-k} & = & - v^{12}_{k}\, a_k - v^{22}_{k}\, b_k + u^{12}_{k}\, a^+_{-k} +  u^{22}_{-k}\, b^+_{-k}. \nonumber
\eea
In Eqs.(\ref{LMJT-INVtrans}) we have used that Bogoliubov coefficients are real and even functions of the wave vector $\bf k$.

Farther on we replace in the equations
\be\label{LMJT-eq1}
\left [\alpha_k , h_q\right ]\, =\,  E^{\alpha}_k\, \alpha_k \qquad \left [\beta_k , h_q\right ]\, =\, E^{\beta}_k\, \beta_k\,\,,
\ee
obtained from Eq.(\ref{LMJTh5}), $\alpha$ and $\beta$ operators with $a$ and $b$ ones and use the equalities
\bea\label{LMJT-eq2}
\left [ a_k , h_q \right ] & = & \varepsilon^a a_k\,-\,2\varepsilon^a_k a^+_k\,-\,\gamma b_k \\
\left [ b_k , h_q \right ] & = & \varepsilon^b b_k\,-\,2\varepsilon^b_k b^+_k\,-\,\gamma a_k\,, \nonumber
\eea
which follow from Eq.(\ref{LMJT-h4}).
Comparing the coefficients in the front of the $a$ and $b$ operators one obtains two systems of equations for the Bogoliubov coefficients:
\bea\label{LMJT-eq3}
\left (E^{\alpha}_k\,-\,\varepsilon^a\right )u^{11}_k\,+\, \gamma u^{21}_k\,+\,2\gamma^a_k v^{11}_k\,=\,0 \nonumber \\
\left (E^{\alpha}_k\,-\,\varepsilon^b\right )u^{21}_k\,+\, \gamma u^{11}_k\,+\,2\gamma^b_k v^{21}_k\,=\,0 \nonumber \\
\left (E^{\alpha}_k\,+\,\varepsilon^a\right )v^{11}_k\,-\, \gamma v^{21}_k\,-\,2\gamma^a_k u^{11}_k\,=\,0 \nonumber \\
\left (E^{\alpha}_k\,+\,\varepsilon^b\right )v^{21}_k\,-\, \gamma v^{11}_k\,-\,2\gamma^b_k u^{21}_k\,=\,0
\eea
and
\bea\label{LMJT-eq4}
\left (E^{\beta}_k\,-\,\varepsilon^a\right )u^{12}_k\,+\, \gamma u^{22}_k\,+\,2\gamma^a_k v^{12}_k\,=\,0 \nonumber \\
\left (E^{\beta}_k\,-\,\varepsilon^b\right )u^{22}_k\,+\, \gamma u^{12}_k\,+\,2\gamma^b_k v^{22}_k\,=\,0 \nonumber \\
\left (E^{\beta}_k\,+\,\varepsilon^a\right )v^{12}_k\,-\, \gamma v^{22}_k\,-\,2\gamma^a_k u^{12}_k\,=\,0 \nonumber \\
\left (E^{\beta}_k\,+\,\varepsilon^b\right )v^{22}_k\,-\, \gamma v^{12}_k\,-\,2\gamma^b_k u^{22}_k\,=\,0
\eea
We supplement the system of equations (\ref{LMJT-eq3}),(\ref{LMJT-eq4}) with two equations which are consequence of the Bose commutation relations\,\,
$[\alpha_k,\alpha^+_k]\,=\,1$ and $[\beta_k,\beta^+_k]\,=\,1$:
\be\label{LMJT-eq3a}
(u^{11}_k)^2\,+\,(u^{21}_k)^2-(v^{11}_k)^2-(v^{21}_k)^2\,=\,1
\ee
\be\label{LMJT-eq4a}
(u^{12}_k)^2\,+\,(u^{22}_k)^2-(v^{12}_k)^2-(v^{22}_k)^2\,=\,1
\ee

 Looking for the solution of the system of equations (\ref{LMJT-eq3}) one arrives at an equation for the dispersion $E^{\alpha}_k$, which is the same as the equation for the dispersion $E^{\beta}_k$ obtained from the system (\ref{LMJT-eq4})
\bea\label{LMJT-eq5}
& & E^4_k-E^2_k [(\varepsilon^a)^2+(\varepsilon^b)^2-2\gamma^2+4(\gamma^a_k)^2+4(\gamma^b_k)^2] \nonumber \\
& + & (\varepsilon^a\varepsilon^b)^2-
2\gamma^2\varepsilon^a\varepsilon^b-4(\varepsilon^a\gamma^b_k)^2-4(\varepsilon^b\gamma^a_k)^2 \nonumber \\
& + & (\gamma^2-4\gamma^a_k\gamma^b_k)^2=0
\eea
The positive solutions of the equation (\ref{LMJT-eq5}) have the form
\be  \label{LMJT_E2}
E^{\pm}_k\,=\,\sqrt{\frac
12\,\left (A_k+B_k\pm\sqrt{(A_k-B_k)^2+4 D_k} \right)},\ee with $A_k,\,B_k$ and $D_k$ given by equations (\ref{LMJTE2}).
For definiteness one sets $E^-_k=E^{\alpha}_k$ and $E^+_k=E^{\beta}_k$.

To present the Bogoliubov coefficients, which are the solutions of the systems of equations (\ref{LMJT-eq3},\ref{LMJT-eq3a}) and (\ref{LMJT-eq4},\ref{LMJT-eq4a}), we introduce the functions
\bea\label{LMJT-M}
M^1_k & = & 2\gamma \left [\gamma^b_k(E^{\alpha}_k+\varepsilon^a)-\gamma^a_k(E^{\alpha}_k-\varepsilon^b)\right ]\nonumber \\
M^2_k & = & 2\gamma^2\gamma^a_k-2\gamma^b_k(E^{\alpha}_k-\varepsilon^a)(E^{\alpha}_k+\varepsilon^a)-8\gamma^b_k(\gamma^a_k)^2 \nonumber\\
M^3_k & = & \gamma (E^{\alpha}_k-\varepsilon^a)(E^{\alpha}_k-\varepsilon^b)-\gamma^3 + 4\gamma\gamma^a_k\gamma^b_k \nonumber \\
M^4_k & = & (E^{\alpha}_k-\varepsilon^a)(E^{\alpha}_k-\varepsilon^b)(E^{\alpha}_k+\varepsilon^a)-\gamma^2(E^{\alpha}_k+\varepsilon^a)\nonumber \\
& & +  4\,(\gamma^a_k)^2(E^{\alpha}_k-\varepsilon^b)
\eea
and
\bea\label{LMJT-T}
R^1_k & = &  2\gamma^2\gamma^b_k-2\gamma^a_k(E^{\beta}_k-\varepsilon^b)(E^{\beta}_k+\varepsilon^b)-8\gamma^a_k(\gamma^b_k)^2 \nonumber\\
R^2_k & = & 2\gamma \left [\gamma^a_k(E^{\beta}_k+\varepsilon^b)-\gamma^b_k(E^{\beta}_k-\varepsilon^a)\right ]\nonumber \\
R^3_k & = & (E^{\beta}_k-\varepsilon^a)(E^{\beta}_k-\varepsilon^b)(E^{\beta}_k+\varepsilon^b)-\gamma^2(E^{\beta}_k+\varepsilon^b)\nonumber \\
& & +  4\,(\gamma^b_k)^2(E^{\beta}_k-\varepsilon^a) \\
R^4_k & = & \gamma (E^{\beta}_k-\varepsilon^a)(E^{\beta}_k-\varepsilon^b)-\gamma^3 + 4\gamma\gamma^a_k\gamma^b_k \nonumber
\eea
In terms of the above functions the expressions for the coefficients are simple:
\bea\label{Bogoliubov1}
u^{11}_k & = & \frac {M^1_k}{\sqrt{(M^1_k)^2+(M^2_k)^2-(M^3_k)^2-(M^4_k)^2}} \nonumber\\
\\
u^{21}_k & = & \frac {M^2_k}{\sqrt{(M^1_k)^2+(M^2_k)^2-(M^3_k)^2-(M^4_k)^2}} \nonumber \\
v^{11}_k & = & \frac {M^3_k}{\sqrt{(M^1_k)^2+(M^2_k)^2-(M^3_k)^2-(M^4_k)^2}} \nonumber \\
v^{21}_k & = & \frac {M^4_k}{\sqrt{(M^1_k)^2+(M^2_k)^2-(M^3_k)^2-(M^4_k)^2}} \nonumber
\eea
\bea\label{Bogoliubov2}
u^{12}_k & = & \frac {R^1_k}{\sqrt{(R^1_k)^2+(R^2_k)^2-(R^3_k)^2-(R^4_k)^2}} \nonumber\\
\\
u^{22}_k & = & \frac {R^2_k}{\sqrt{(R^1_k)^2+(R^2_k)^2-(R^3_k)^2-(R^4_k)^2}} \nonumber \\
v^{12}_k & = & \frac {R^3_k}{\sqrt{(R^1_k)^2+(R^2_k)^2-(R^3_k)^2-(R^4_k)^2}} \nonumber \\
v^{22}_k & = & \frac {R^4_k}{\sqrt{(R^1_k)^2+(R^2_k)^2-(R^3_k)^2-(R^4_k)^2}} \nonumber
\eea

Finally, one can represent the vacuum energy Eq.(\ref{LMJTh5}) in the form
\begin{widetext}\bea\label{LMJT-vE}
E^0_k & = & \frac 12 \left [E^{\alpha}_k+ E^{\beta}_k -\varepsilon^a-\varepsilon^b \right] \nonumber \\
 & + &  \frac {\gamma M^1_kM^2_k-\gamma M^3_kM^4_k + \frac 14 (\varepsilon^b-\varepsilon^a)[(M^1_k)^2-(M^2_k)^2-(M^3_k)^2+(M^4_k)^2]}{(M^1_k)^2+(M^2_k)^2-(M^3_k)^2-(M^4_k)^2} \nonumber\\
 \\
& + & \frac {\gamma R^1_kR^2_k-\gamma R^3_kR^4_k + \frac 14 (\varepsilon^b-\varepsilon^a)[(R^1_k)^2-(R^2_k)^2-(R^3_k)^2+(R^4_k)^2]}{(R^1_k)^2+(R^2_k)^2-(R^3_k)^2-(R^4_k)^2}\nonumber  \eea\end{widetext}

\emph{Supplemental material B}

For common discussion of the spin-wave excitations in $\textbf{A}$, $\textbf{C}$ and $\textbf{G}$ antiferromagnetic phases it is convenient to consider a theory with Hamiltonian
\be\label{ACG-1}
h=\sum\limits_{i \mu} J^{\mu} {\bf S}_i \cdot {\bf S}_{i+{\bf e}_{\mu}}, \ee
where ${\bf e}_{\mu}$ is the unit vector along $\mu=x,y,z$ and $J^{\mu}$ is the exchange constant which depends on the space directions ($J^x,J^y,J^z$). To obtain the ground state magnetic order we represent the spin operators as vectors ${\bf S}_i=s{\bf n}_i$, where ${\bf n}_i$ is an unit vector in the form ${\bf n}_i=(\sin\theta_i,0,\cos\theta_i)$.
We consider a simplest dependence of the angle $\theta_i$  on the lattice site $\theta_i={\bf r}_i\cdot {\bf Q}$, where ${\bf Q}=(Q_x,Q_y,Q_z)$. The ground state energy, obtained from the Hamiltonian Eq.\eqref{ACG-1} is
\be\label{ACG-2}
h_{gr}=s^2 N \sum\limits_{\mu} J^{\mu}\cos Q_{\mu}, \ee
where $N$ is the number of the lattice sites. The physical value of the wave vector ${\bf Q}$ is the value at which the ground state energy $h_{gr}$ is minimal.

In the case when all three parameters are positive $(J^x>0,J^y>0,J^z>0)$ the physical wave vector is ${\bf Q}={\bf Q^G}=(\pi,\pi,\pi)$ and spin vectors, on nearest neighbor sites, are anti-aligned so that the net magnetization is zero. The state is said to be \textbf{G}-type antiferromagnetic.

When $J^x>0,J^y>0$ and $J^z<0$ the ground state energy is minimal at ${\bf Q}={\bf Q}^C=(\pi,\pi,0)$. The spins are anti-aligned in $x-y$ plane, and parallel along the $z$ direction. This state is a \textbf{C}-type antiferromagnetic state.

Finally, when  $J^x<0,J^y<0$ and $J^z>0$ the physical wave vector is ${\bf Q}={\bf Q}^A=(0,0,\pi)$. The spins are parallel in  $x-y$ plane, and antiparallel along the $z$ direction. This phase is known as \textbf{A}-type antiferromagnetism.

In all three cases $\sin \theta_i=0$. Therefor we can use the representation \eqref{LMJT-HP} for the spin operators. In terms of the bose operators ($a^+_i,a_i$) the Hamiltonian \eqref{ACG-1} reads
\bea\label{ACG-3}
& & h= \sum\limits_{i \mu} J^{\mu}\left [\cos Q_{\mu} \left (s-a^+_i a_i\right) \left (s-a^+_{i+e_{\mu}} a_{i+e_{\mu}} \right)\right.\nonumber \\
& & \left. + \frac 12 \cos\frac {Q_{\mu}}{2}\left( f_i a_i a^+_{i+e_{\mu}} f_{i+e_{\mu}}+a^+_i f_i f_{i+e_{\mu}}a_{i+e_{\mu}}\right) \right. \\
& & \left. - \frac 12 \sin \frac {Q_{\mu}}{2} \left(f_i a_i f_{i+e_{\mu}}a_{i+e_{\mu}}+a^+_if_i a^+_{i+e_{\mu}} f_{i+e_{\mu}}\right)\right]\nonumber . \eea
where $f_i=\sqrt{2s-a^+_i a_i}$.

In the spin-wave approximation $f_i\approx \sqrt{2s}$ and we keep only quadratic terms of the Bose fields $(a^+_i a_i)$
\bea\label{ACG-4}
h_{sw} & = & \sum\limits_{i \mu} J^{\mu}\left [-s\cos Q_{\mu} \left (a^+_i a_i+a^+_{i+e_{\mu}} a_{i+e_{\mu}} \right)\right.\nonumber \\
& & \left. + s \cos\frac {Q_{\mu}}{2}\left(a_i a^+_{i+e_{\mu}}+a^+_i a_{i+e_{\mu}}\right) \right. \\
& & \left. - s \sin \frac {Q_{\mu}}{2} \left( a_i a_{i+e_{\mu}}+a^+_i a^+_{i+e_{\mu}} \right)\right]\nonumber . \eea
For \textbf{G}-type antiferromagnetic systems $({\bf Q}={\bf Q^G})$ the Hamiltonian is
\be\label{ACG-5}
h^G_{sw} =  s \sum\limits_{i \mu} J^{\mu}\left [a^+_i a_i+a^+_{i+e_{\mu}} a_{i+e_{\mu}}
 -  a_i a_{i+e_{\mu}}-a^+_i a^+_{i+e_{\mu}}\right]. \ee
For \textbf{C}-type antiferromagnetic systems $({\bf Q}={\bf Q^C})$ it is
\bea\label{ACG-6}
& & h^C_{sw} =  \sum\limits_{i}\left[\sum\limits_{\mu=x,y} s J^{\mu} \left (a^+_i a_i+a^+_{i+e_{\mu}} a_{i+e_{\mu}} \right)\right. \\
& & \left. +  s J^z \left(a_i a^+_{i+e_{z}}+a^+_i a_{i+e_{z}}-a^+_i a_i-a^+_{i+e_{z}} a_{i+e_{z}}  \right) \right.\nonumber \\
& & \left. - \sum\limits_{\mu=x,y} s J^{\mu}\left( a_i a_{i+e_{\mu}}+a^+_i a^+_{i+e_{\mu}} \right)\right]\nonumber . \eea
Finally, the Hamiltonian of the \textbf{A}-type antiferromagnetic systems $({\bf Q}={\bf Q^A})$ is
\bea\label{ACG-7}
& & h^A_{sw} = \sum\limits_{i}\left[ sJ^{z} \left( a^+_i a_i+a^+_{i+e_{z}} a_{i+e_{z}}- a_i a_{i+e_{z}}-a^+_i a^+_{i+e_{z}} \right)\right.\nonumber \\
\\
& & \left. + \sum\limits_{\mu=x,y} s J^{\mu} \left (a_i a^+_{i+e_{\mu}}+a^+_i a_{i+e_{\mu}}- a^+_i a_i-a^+_{i+e_{\mu}} a_{i+e_{\mu}}\right) \right]\nonumber
\eea

In momentum space representation the Hamiltonians Eqs.\eqref{ACG-5},\eqref{ACG-6} and \eqref{ACG-7} have the form
\be \label{ACG-8}
 h_{sw} = \sum\limits_{k\in B}\left [\varepsilon_k\,a_k^+a_k\, - \, \gamma_k(a^+_ka^+_{-k}+a_{-k}a_k)\right ],
\ee
where for \textbf{G}-type $(J^x>0,J^y>0,J^z>0)$
\bea\label{ACG-9}
\varepsilon_k^G & = & 2s\left( J^x\,+\,J^y\,+\,J^z\right)  \\
\gamma_k^G & = & s \left(J^x \,\cos k_x\,+J^y\,\cos k_y\, +\, J^z\, \cos k_z\right) \nonumber,
\eea
for \textbf{C}-type $(J^x>0,J^y>0,J^z<0)$
\bea\label{ACG-10}
\varepsilon_k^A & = & 2s\left( J^x\,+\,J^y\right)\,+\,2s|J^z|\,\left(1-\cos k_z\right) \nonumber \\
\gamma_k^C & = & s \left(J^x \,\cos k_x\,+J^y\,\cos k_y\right)
\eea
and for  \textbf{A}-type $(J^x<0,J^y<0,J^z>0)$
\bea\label{ACG-11}
\varepsilon_k^A & = & 2s\left(|J^x|\left[1-\cos k_x\right]+|J^y|\left[1-\cos k_y\right]+J^z\right)\nonumber \\
\gamma_k^C & = & s\,J^z \,\cos k_z
\eea

Next, we diagonalize the Hamiltonian Eq.\eqref{ACG-8} by means of the Bogoliubov transformation. In terms of the Bogoliubov operators the Hamiltonian is
\be\label{ACG-12}
h_{sw} = \sum\limits_{k\in B}\left [E_k\,\alpha_k^+\alpha_k\,+\,E^0_k \right ],
\ee
with energy of the Bogoliubov excitations
\be\label{ACG-13}
E_k\,=\,\sqrt{\varepsilon_k^2\, - \, 4\gamma_k^2}.
\ee

The energy of the \textbf{G}-type antiferromagnetic system $E^G_k=\sqrt{(\varepsilon_k^G)^2\, - \, 4(\gamma_k^G)^2}$ is zero at wave vectors ${\bf k}=(0,0,0)$ and ${\bf k} ={\bf Q^G}$. Near these wave vectors the dispersion is linear
\bea\label{ACG-14}
& & E^G_{ {\bf k}\rightarrow {\bf 0}}\simeq 2s\sqrt{J^x+J^y+J^z}\sqrt{\sum\limits_{\mu}J^{\mu} k_{\mu}^2} \\
& & E^G_{{\bf k}\rightarrow {\bf Q^G}}\simeq 2s\sqrt{J^x+J^y+J^z}\sqrt{\sum\limits_{\mu}J^{\mu} (k_{\mu}-Q^G_{\mu})^2}.\nonumber
\eea
The energy $E^C_k=\sqrt{(\varepsilon_k^C)^2\, - \, 4(\gamma_k^C)^2}$ is zero at wave vectors ${\bf k}=(0,0,0)$ and ${\bf k} ={\bf Q^C}$. Near these wave vectors the dispersion is linear
\bea\label{ACG-15}
& & E^C_{ {\bf k}\rightarrow {\bf 0}}\simeq 2s\sqrt{J^x+J^y}\sqrt{\sum\limits_{\mu}|J|^{\mu} k_{\mu}^2} \\
& & E^C_{{\bf k}\rightarrow {\bf Q^C}}\simeq 2s\sqrt{J^x+J^y}\sqrt{\sum\limits_{\mu}|J|^{\mu} (k_{\mu}-Q^C_{\mu})^2}.\nonumber
\eea
Finally, the energy $E^A_k=\sqrt{(\varepsilon_k^A)^2\, - \, 4(\gamma_k^A)^2}$ is zero at wave vectors ${\bf k}=(0,0,0)$ and ${\bf k} ={\bf Q^A}$. Near these wave vectors the dispersion is linear
\bea\label{ACG-16}
& & E^A_{ {\bf k}\rightarrow {\bf 0}}\simeq 2s\sqrt{J^z}\sqrt{\sum\limits_{\mu}|J|^{\mu} k_{\mu}^2} \\
& & E^A_{{\bf k}\rightarrow {\bf Q^A}}\simeq 2s\sqrt{J^z}\sqrt{\sum\limits_{\mu}|J|^{\mu} (k_{\mu}-Q^A_{\mu})^2}\nonumber
\eea

The dispersions \eqref{ACG-14},\eqref{ACG-15} and \eqref{ACG-16} allow to conclude that magnon fluctuations in \textbf{A}-type, \textbf{C}-type and \textbf{G}-type antiferromagnets are identical (two Goldstone bosons with linear dispersion). This is exactly what we claim in the Summary section of the present paper.


\begin{thebibliography}{99}
\bibitem{LMJTnote} The vertexes of the Brillouin zone are topologically equivalent. This is why we have two long-range excitations in the
theory.
\end{thebibliography}
\end{document}